\documentclass[12pt]{iopart} 
\usepackage{amsfonts,amssymb}
\usepackage{iopams,cite} 
\usepackage{graphicx} 
\usepackage{color}

\newcommand{\text}[1]{\mbox{\scriptsize{#1}}}

\def\rd{{\rm d}}

\def\dbar{{\mathchar'26\mkern-12mu \rd}}

\begin{document}

\title[Entropy Production]{Entropy Production in Mesoscopic Stochastic Thermodynamics: Nonequilibrium Kinetic Cycles Driven by Chemical Potentials, Temperatures, and Mechanical Forces}
\author{Hong Qian,$^{1}$ Signe Kjelstrup,$^{2}$ Anatoly B. Kolomeisky$^{3}$ and Dick Bedeaux$^{2}$}

\address{$^{1}$  University of Washington, Department of Applied Mathematics, Seattle, WA 98195, USA}

\address{$^{2}$  Norwegian University of Science and Technology, Department of Chemistry, Trondheim,NO-7491, Norway}

\address{$^{3}$ Rice University, Department of Chemistry and Center for Theoretical Biological Physics, 6100 Main Street, Houston, TX 77005-1892, USA}

\eads{\mailto{tolya@rice.edu}}

\begin{abstract}

Nonequilibrium thermodynamics (NET) investigates  processes in systems out of global equilibrium. On a mesoscopic level, it provides a statistical dynamic description of various complex phenomena such as chemical reactions, ion transport, diffusion, thermochemical, thermomechanical and mechanochemical fluxes.  In the present review, we introduce a mesoscopic stochastic formulation of NET by analyzing entropy production in several simple examples. The fundamental role of nonequilibrium steady-state cycle kinetics is emphasized. The statistical mechanics of Onsager's reciprocal relations in this context is elucidated. Chemomechanical, thermomechanical, and enzyme-catalyzed thermochemical energy transduction processes are discussed. It is argued that mesoscopic stochastic NET provides a rigorous mathematical basis of fundamental concepts needed for understanding complex processes in chemistry, physics and biology, and which is also relevant for nanoscale technological advances.

\end{abstract}


\maketitle

\section{Introduction}

Nonequilibrium thermodynamics (NET) concerns with dynamic processes in systems that are not in global equilibrium, either in a transient or in a stationary state. Since only few  systems can be viewed as really equilibrium, the subject has a fundamental importance for understanding various phenomena in Chemistry, Physics and Biology. It has a long history, starting with famous studies of Thomson on thermoelectricity  \cite{Thomson1856}. The work of Onsager \cite{onsager-31, onsager-31b} has laid the foundation of the field; it puts the earlier research by Thomson, Boltzmann, Nernst, Duhem, Jauman and Einstein into a systematic framework. By following Onsager, a consistent NET of continuous systems was developed in the 1940s by Meixner \cite{Meixner41,Meixner43}, and Prigogine \cite{prigogine}. Many key aspects of the Onsager's theory were clarified by Casimir \cite{casimir-45}. The most general description of NET, so far, is the well-known  book by de Groot and Mazur from 1962 \cite{degroot-mazur}.

The basic principles of thermodynamics asserts the existence of a special function of the macroscopic state of the system, which is called entropy $S$. This entropy satisfies the following balance equation:\footnote{In classical thermodynamics, a distinction between the total differential of a quantity $Q$, $\mathrm{d}Q$, and an inexact differential $\dbar Q$, which is path dependent, has to be explicitly made.}
\begin{equation}\label{1}
\frac{\mathrm{d}S}{\mathrm{d}t}=\frac{\dbar_{e}S}{\mathrm{d}t}+\frac{\dbar_{i}S}{\mathrm{d}t},  
\end{equation}
in which $\dbar_{e}S/\mathrm{d}t$ is the entropy supplied by the system's environment, and $\dbar_{i}S/\mathrm{d}t$ is the always non-negative entropy production inside the system. The sign of $\dbar_{e}S/\mathrm{d}t$, however, can be positive, zero or negative. For an isolated system that has no entropy exchange with its environment, $S$ always increases until it attains the maximum. The system then reaches the equilibrium.

\section{Theories of Nonequilibrium Thermodynamics}

 Eq. (\ref{1}) is a fundamental relation that describes the entropy production. It plays a key central role in NET. Macroscopic NET as presented in \cite{degroot-mazur} treats various processes in the absence of fluctuations. This will be discussed shortly in Subsection 2.1. As explained by Ortiz de Z\'{a}rate and Sengers \cite{ortiz2006}, it is possible to extend nonequilibrium thermodynamics to include hydrodynamic fluctuations in driven systems using appropriate fluctuation-dissipation theorems. We will not go into this direction, but rather describe a novel {\it mesoscopic} NET with fluctuations in phase space in terms of time-dependent and stationary probability distributions. The main focus of this paper is to present a theoretical framework that will show how thermodynamic forces and fluxes in various realistic nonequilibrium processes can all be represented in terms of a unified treatment at the mesoscopic level, in phase space.

\subsection{Macroscopic nonequilibrium thermodynamics}

\label{sec:2.1}

There are several theories for nonequilibrium systems that start with the entropy balance equation. De Groot and Mazur's approach \cite{degroot-mazur}, followed by Kjelstrup and Bedeaux \cite{bedeaux-kjelstrup} for heterogeneous systems, obtained a spatially resolved version of Eq. (\ref{1}). For a homogeneous fluid in terms of continuous densities it can be rewritten as
\begin{eqnarray}
S(t) &=&\int_{V}s(x,t)\mathrm{d}V,  \label{2} \\
\frac{\dbar_{e}S}{\mathrm{d}t}
&=&-\int_{\partial V}\mathbf{J}_{s}(x,t)\cdot \mathrm{d}S  \label{3} \\
\frac{\dbar_{i}S}{\mathrm{d}t}
&=&\int_{V}\sigma (x,t)\mathrm{d}V,  \label{4}
\end{eqnarray}
where $s(x,t)$ is the entropy density per unit volume, $\mathbf{J}_{s}(x,t)$  is the spatial entropy flux, and $\sigma (x,t)$ is the entropy production per unit volume, which is non-negative. The second integral is over the surface of the volume $V$, and $\mathrm{d}S$ has the direction of the outward normal. Applying  Gauss' theorem to Eq. (\ref{1})
in an arbitrary sub-volume, one obtains
\begin{equation}
\frac{\partial s(x,t)}{\partial t}=-\mbox{div}\mathbf{J}_{s}(x,t)+\sigma(x,t).  \label{5}
\end{equation}
In order to calculate $\partial s(x,t)/\partial t$  one follows the Gibbs equation, 
\begin{equation}
\mathrm{d}u=T\mathrm{d}s+\sum_{j=1}^{n}\mu _{j}\mathrm{d}c_{j},  \label{6}
\end{equation}
where $u,T,\mu _{j},c_{j}$\ are the internal energy density, temperature and the chemical potentials and molar densities of component $j$, respectively. The use of relation (\ref{6}) implies the assumption of { \it local equilibrium} in space and time, meaning that all thermodynamic relations remain valid at a coarse-grained scale that is {\it macroscopically small but microscopically large}.

The balance equations for the internal energy and the component densities are 
\begin{eqnarray}
\frac{\partial u}{\partial t} &=&-\mbox{div}\mathbf{J}_{u}\text{ },
\label{7} \\
\frac{\partial c_{j}}{\partial t} &=&-\mbox{div}\big(c_{j}\mathbf{v}_{j}\big)%
+\sum_{\ell =1}^{m}\nu _{\ell j}r_{\ell },  \label{8}
\end{eqnarray}
where $\mathbf{J}_{u}$\ is the internal energy flux, $c_{j}\mathbf{v}_{j}$ are the spatial molar fluxes, $r_{\ell }$ is the rate of the $\ell ^{th}$ chemical reaction, and $\nu _{\ell j}$\ are the corresponding stoichiometric coefficients. The internal energy flux $\mathbf{J}_{u}$ and the velocities $\mathbf{v}_{j}$ are in the laboratory frame of reference. We consider no external potentials and restrict ourselves to mechanical equilibrium. Furthermore we neglect viscous contributions to the pressure. Substituting Eqs. (\ref{7}) and (\ref{8}) into Eq. (\ref{6}) results in
\begin{eqnarray}
\frac{\partial s}{\partial t} &=&-\mbox{div}\left( \frac{\mathbf{J}_{u}-\sum_{j=1}^{n}\mu _{j}c_{j}\mathbf{v}_{j}}{T}\right) +\mathbf{J}_{u}\cdot \mbox{grad}\left( \frac{1}{T}\right)  \nonumber \\
&&-\sum_{j=1}^{n}c_{j}\mathbf{v}_{j}\cdot \mbox{grad}\left( \frac{\mu _{j}}{T}\right) -\sum_{\ell =1}^{m}r_{\ell }\left( \frac{\Delta G_{\ell }}{T}\right),  \label{9}
\end{eqnarray}
where $\Delta G_{\ell }\equiv \sum_{j=1}^{n}\nu _{\ell j}\mu _{j}$ is the Gibbs energy difference of the $\ell ^{th}$ chemical reaction. Comparing with Eq. (\ref{5}) yields the entropy flux and the entropy production:
\begin{eqnarray}
\mathbf{J}_{s} &=&\frac{\mathbf{J}_{u}}{T}-\frac{1}{T}\sum_{j=1}^{n}\mu_{j}c_{j}\mathbf{v}_{j},  \label{10} \\
\sigma &=&\mathbf{J}_{u}\cdot \mbox{grad}\left( \frac{1}{T}\right)-\sum_{j=1}^{n}c_{j}\mathbf{v}_{j}\cdot \mbox{grad}\left( \frac{\mu _{j}}{T}\right) -\sum_{\ell =1}^{m}r_{\ell }\left( \frac{\Delta G_{\ell }}{T}\right).  \label{11}
\end{eqnarray}

While de Groot and Mazur \cite{degroot-mazur}\ always used fluxes in the barycentric (center-of-mass) frame of reference, we utilized here the fluxes in the laboratory frame of reference. The total entropy production has the important property of being invariant under the transformation of one frame of reference to another. This may easily be verified by defining the heat flux $\mathbf{J}_{q}=\mathbf{J}_{u}-u\mathbf{v}$ and the diffusion fluxes $\mathbf{J}_{j}=c_{j}\left( \mathbf{v}_{j}-\mathbf{v}\right) $, where the velocity $\mathbf{v}$ can be chosen to be the barycentric velocity, the mean molar velocity, the mean volume velocity, the velocity of one of the components (the solvent), or the velocity of the surface of for instance an electrode. We refer to Chapter 11 of \cite{degroot-mazur} for a precise definition of these velocities and a detailed discussion. Regarding the use of a surface as the frame of reference we refer to Kjelstrup and Bedeaux  \cite{bedeaux-kjelstrup}. Substituting these definitions\ into Eq. (\ref{11}), and with Gibbs-Duhem relation and mechanical equilibrium, it follows that
\begin{equation}
\sigma =\mathbf{J}_{q}\cdot \mbox{grad}\left( \frac{1}{T}\right)-\sum_{j=1}^{n}\mathbf{J}_{j}\cdot \mbox{grad}\left( \frac{\mu _{j}}{T}\right) -\sum_{\ell =1}^{m}r_{\ell }\left( \frac{\Delta G_{\ell }}{T}\right).  \label{12}
\end{equation}

The entropy production is a binary product of so-called conjugate thermodynamic fluxes and forces. For different choices of $\mathbf{v}$, the heat flux and the diffusion fluxes are different, and they can be chosen depending on the experimental setting. When one introduces alternative thermodynamic fluxes one should realize that the corresponding conjugate thermodynamic forces may also change. An example is the use of the measurable heat flux \cite{bk-08}: 
\begin{equation}
\mathbf{J}_{q}^{\prime }\equiv \mathbf{J}_{q}-\sum_{j=1}^{n}h_{j}\mathbf{J}_{j},  \label{13}
\end{equation}
in which $h_{j}$ is the enthalpic contribution to $\mu _{j}$, $\mu_{j}=h_{j}-Ts_{j}$. When we substitute of this definition in Eq. (\ref{12}) and use van't Hoff's equation $h_{j}=\partial (\mu _{j}/T)/\partial (1/T)$ \cite{qian-jjh,qian-eec-98}, the entropy production becomes
\begin{equation}
\sigma =\mathbf{J}_{q}^{\prime }\cdot \mbox{grad}\left( \frac{1}{T}\right) -\frac{1}{T}\sum_{j=1}^{n}\mathbf{J}_{j}\cdot \big(\mbox{grad}\mu _{j}\big)_{T}-\sum_{\ell =1}^{n}r_{\ell }\left( \frac{\Delta G_{\ell }}{T}\right),
\label{14}
\end{equation}
The subscript $T$ in the $(\mbox{grad}\mu _{j})_{T}$ means that the spatial differentiation is calculated keeping the temperature constant. One can further show that the measurable heat flux is independent of the frame of reference \cite{degroot-mazur,bedeaux-kjelstrup}. Therefore, this is the heat flux which is most convenient for the interpretation of experiments. Using the conjugate fluxes and forces in, for instance, Eq. (\ref{14}) one can express the vectorial fluxes $\{\mathbf{J}_{q}^{\prime },\mathbf{J}_{j}\}$ linearly in the vectorial forces $\big\{\mbox{grad}T^{-1},-T^{-1}(\mbox{grad}\mu_{k})_{T}\big\}$. The proportionality matrix was shown to be symmetric by Onsager \cite{onsager-31,onsager-31b} using microscopic reversibility. In a 3D isotropic system such as a fluid the vectorial fluxes do not couple to the scalar forces driving the chemical reactions according to Curie's principle. The net reaction fluxes are proportional to $-\Delta G_{\ell}/T$ in the linear description. When one considers transport into and through surfaces \cite{bedeaux-kjelstrup} the fluxes normal to the surface are also scalars. As a consequence, a chemical potential difference across the surface may drive a chemical reaction at the surface (membrane).

The entropy production is non-negative according to the second law. As the vectorial and the scalar contributions do not couple according to the Curie's principle, one may show that the total entropy production due to the vectorial contributions and the entropy production due to the scalar contribution are both positive. Neither the separate vectorial nor the
separate scalar terms have to be positive. Energy transduction occurs when a larger positive term overcomes a smaller negative term \cite{hill-book}.

The vectorial contributions are zero when a chemical system is rapidly stirred. For multiple reactions the above 3D theory can then be reduced to Qian and Beard's stoichiometric network theory, which has found a successful application in metabolic engineering \cite{qian-beard-05}. See \cite{ge-qian-2013} for an extensive discussion on nonequilibrium steady states with regenerating system and quasi-steady state with excess chemicals using buffers and chelators. At this level, the type of ensemble, and what are controlled thermodynamic variables, matters. This is a very important result that was first discussed by Hill \cite{hill-nanobook}.

\subsection{Stochastic Liouville dynamics}

\label{sec:2.2}

As a point of departure from the macroscopic NET theory presented above, mesoscopic NET is based on a conservation law in the phase space of any dynamics: the Chapman-Kolmogorov equation for the conservation of probability in equations of motions in the broadest sense.\footnote{For deterministic, Hamiltonian systems, the dynamics of probability is formulated in term of a measure-theoretical transfer operator, also known as Ruelle-Perron-Frobenius operator \cite{gaspard}.} Instead of being based on the entropy balance Eq. (\ref{1}), the mesoscopic NET derives a mesoscopic version of it with a dynamic foundation \cite{mackey}, together with an explicit expression for the probability flux, and proving the
non-negativity of entropy production. As will be shown in Sec. \ref{sec:3}, the probability flux in phase space can be interpreted, based on a {\it local equilibrium} assumption, to the laboratory measurements of various realistic fluxes, such as chemical reaction flux, heat, mass transport, electrical, etc.

Such a mesoscopic theory of NET, in terms of a stochastic description of dynamics in phase space, has been repeatedly alluded to by many scientists. An earlier reference is the theory of stochastic Liouville dynamics \cite{bergmann-lebowitz,lebowitz-bergmann}. We shall not present this theory in detail. Instead, we discuss the logic relation of this work to the classical work of Boltzmann and others. The present work will then focus on overdamped stochastic dynamics, which is valid for studying NET of soft condensed matter, solution chemistry and biochemistry. The stochastic description gives a natural extension of NET to mesoscopic systems, which contains fluctuations.

Statistical or kinetic theories of nonequilibrium phenomena, formulated in terms of measurable quantities in 3D physical space, such as the Boltzmann equation, provide a more detailed mechanism for dynamic processes \cite{degroot-mazur}. Such theories have, however, only been developed for special classes of phenomena and use particular molecular models. Although going deeper into the physical description, they do not give a general framework for the description of transport processes \cite{degroot-mazur}.

There have been several theories of irreversible phenomena using stochastic processes \cite{vankampen}. The Klein-Kramers equation \cite{klein,kramers} and the Langevin equation, together with fluctuation-dissipation relation, are a natural extension of classical conservative dynamics of a Hamiltonian system in contact with a heat bath. Cox \cite{cox-1,cox-2} developed a Markov theory for irreversible processes that generalized Gibbs statistical mechanics to irreversible processes. Cox's work was motivated by the consideration that \textquotedblleft In the theory of time-dependent thermal
phenomena, the method of Gibbs appears to have been rather neglected in comparison with that of Boltzmann.\textquotedblright  \cite{cox-1} Onsager and Machlup developed a comprehensive linear stochastic dynamical theory based on a Gaussian Markov description, e.g., Ornstein-Uhlenbeck processes \cite{onsager-machlup,machlup-onsager}. The Shannon entropy has been introduced naturally in these theories, as the dynamic counterpart of the entropy of Gibbsian statistical ensemble, an insight originated in Boltzmann's kinetic theory and his H-function. None of these works, however, connected the
stochastic dynamics with the entropy balance equation in Eq. (\ref{1}).

By using a Liouville formulation of general conservative dynamics in phase space, together with a stochastic kernel, Bergmann and Lebowitz \cite{bergmann-lebowitz,lebowitz-bergmann} assumed the entropy balance equation (Eq. 1)\footnote{For a given dynamics and a definition of $S$, $dS/dt$ can always be computed. The entropy production as in Eq.(\ref{1}), however, has always been defined phenomenologically based on physical intuitions. This situation has changed since the emergence of a measure-theoretical definition(s) of entropy production in the theory of Markov dynamics \cite{jqq}.} and introduced $\mathrm{d}S_{total}(t)/\mathrm{d}t=\mathrm{d}S(t)/\mathrm{d}t-T^{-1}(\mathrm{d}U/\mathrm{d}t)\geq 0$ as the total entropy change, of the system and the heat bath together. They were able to show that the Helmholtz energy of a closed system, which was expressed in terms of the time-dependent probability density function $f(x,t)$ as 
\begin{equation}
F\big[f]=U-TS=\int_{x}f(x,t)\Big[H(x)+k_{\text{B}}T\ln f(x,t)\Big]\mathrm{d}x,  \label{15}
\end{equation}
was monotonic and non-increasing. Here $H$ is the Hamiltonian and $k_{\text{B}} $ is the Boltzmann's constant. Furthermore, $x$ is a point in the phase-space of the system. A fluctuation-dissipation relation for a stochastic kernel
with temperature $T$ was also obtained for systems that approach to $f^{eq}(x)=\exp [-H(x)/k_{\text{B}}T]$. Finally, they proved that Liouville dynamics with multiple heat-baths at different temperatures yield a nonequilibrium steady state (NESS) of the closed system, with a positive entropy production.

\subsection{Mesoscopic stochastic thermodynamics}

While the stochastic Liouville dynamics discussed in Sec. \ref{sec:2.2}, as a dynamic counterpart to the equilibrium statistical thermodynamics based on a microcanonical ensemble, has the virtue of being rooted in Newtonian mechanics, its applicability to condensed matter chemistry, polymer systems, and biochemistry, is limited. In chemistry  it is the  Gibbsian statistical thermodynamics based on a canonical ensemble that has wide and successful applications. Overdamped stochastic dynamical theory of a polymer solution is an example of such success with many applications \cite{doi-edwards,qian-pre-02}.

This observation motivated a stochastic dynamics formulation of NET in phase space. The approach in Sec. \ref{sec:2.1} assumes the validity of local equilibrium, meaning that all thermodynamic relations are valid locally. A mesoscopic theory can be developed based on a Markovian probabilistic description. The state space can be discrete or continuous. The
Chapman-Kolmogorov equation for a Markov process can then be used to obtain a master equation. For a discrete-state space one has 
\begin{equation}
\frac{\mathrm{d}p_{i}(t)}{\mathrm{d}t}=\sum_{j}\big[J_{ji}(t)-J_{ij}(t)\big]=\sum_{j}\big[p_{j}(t)q_{ji}-p_{i}(t)q_{ij}\big],  \label{eqn-16}
\end{equation}
where $p_{i}(t)$\ is the probability of the system being in state $i$\ at time $t$. Furthermore, $J_{ij}(t)=p_{i}(t)q_{ij}$ is the one-way flux from state $i$ to state $j$ at time $t$. For a continuous-state space the master equation becomes \begin{eqnarray}
\frac{\partial f(x,t)}{\partial t} &=&\int dx^{\prime }\Big[J\left(x^{\prime },x;t\right) -J(x,x^{\prime };t)\Big]  \nonumber \\
&=&\int dx^{\prime }\Big[f(x^{\prime },t)q\left( x^{\prime },x\right)-f(x,t)q(x,x^{\prime })\Big],  \label{m1}
\end{eqnarray}
where $f(x,t)$ is the density of the probability of the system being in state $x$\ at time $t$. Similarly, $J(x,x^{\prime };t)=f(x,t)q(x,x^{\prime})$ is the one-way flux density from state $x$ to state $x^{\prime }$ at time $t$. If a system is not driven, then it reaches equilibrium as its stationary state. In equilibrium it follows from microscopic reversibility that the system satisfies  a detailed balance: \cite{degroot-mazur,vankampen} 
\begin{eqnarray}
p_{j}^{eq}q_{ji} &=&p_{i}^{eq}q_{ij}, \\
f^{eq}(x^{\prime })q\left( x^{\prime },x\right) &=&f^{eq}(x)q(x,x^{\prime }).
\end{eqnarray}
The superscript $eq$\ indicates the equilibrium probability distributions for discrete systems, or probability densities of continuous systems. When the system is not in equilibrium it does not satisfy detailed balance.  Yet, dynamics whose stationary state  possesses detailed balance has a stringent constraint on its rate coefficients; this is known as a Wegscheider condition \cite{gnlewis} and Kolmogorov cycle criterion in the  Markov-process theory.  A system may also, of course,  be driven by constant external force. As a consequence, a stationary state may develop which does not satisfy detailed balance.

\subsubsection{Detailed balance}

At this point, it is important to clearly explain the term ``detailed balance'' because of frequent confusions and wrong applications. As we just stated above, it follows from  a microscopic reversibility that the probabilistic description of a thermodynamic equilibrium system satisfies the detailed balance. For proof we refer to \cite{degroot-mazur,vankampen}. In nature, there are many systems that never come to equilibrium. In a living being, for instance, ions are continuously pumped by ATPases through membranes. Equilibrium is obtained only when the living being dies. It follows that the detailed balance, though exact in equilibrium, is not  relevant for a description of living biological systems. In general, any nonequilibrium state in such open systems is maintained by, for instance, by constantly adding ATP or other reactants. In the description of the behavior of such systems, it is common to introduce pseudo-first-order rate coefficients to replace the original coefficient. This is done by absorbing the probabilities of buffered components, maintained at a constant nonequilibrium value, in the rate constants. The product defines the new pseudo-first-order rate coefficient. The resulting description concerns then the  behavior far from equilibrium, and the forward and backward rates are then evidently not balanced, even in a stationary state. In the  {\it buffered pseudo-first-order rate coefficients} the system does not have an equilibrium state, and therefore never satisfies detailed balance.

Detailed balance is also a mathematical concept in the theory of Markov process and Monte Carlo statistical simulations. The mathematical concept of detailed balance is applicable to Markov models of physical and chemical
origin in closed systems. A Markov model for an open (buffered) nonequilibrium system using the above mentioned pseudo-first-order rate coefficients does not satisfy detailed balance. We refer to \cite{qian-jpc-06,qian_arpc} for a detailed discussion. In the present work, we will also use the pseudo-first-order rate coefficients when this is convenient.

\subsubsection{Entropy balance equation for continuous Markov dynamics}

We consider a transition probability $q(x^{\prime },x)$ from state $x^{\prime }$ to state $x,\ $which is sharply peaked in the sense that $f(x^{\prime },t)$\ varies slowly over the range of $q(x^{\prime },x)$. One may then use a moment expansion of the transition probability to the second order:
\begin{equation}
q(x^{\prime },x)=q_{1}(x^{\prime })\cdot \frac{\partial }{\partial x^{\prime
}}\delta \left( x-x^{\prime }\right) +\frac{1}{2}q_{2}(x^{\prime })\frac{\partial ^{2}}{\partial x^{\prime 2}}\delta \left( x-x^{\prime }\right) .
\label{m2}
\end{equation}
Both $q_{1}$\ and $\frac{\partial }{\partial x^{\prime }}$\ are vectors in phase-space, the period $\cdot $ indicates a contraction, and $\frac{\partial ^{2}}{\partial x^{\prime 2}}\equiv \frac{\partial }{\partial
x^{\prime }}\cdot \frac{\partial }{\partial x^{\prime }}$. A possible zeroth order contribution does not contribute to $\partial f/\partial t$. In the moment expansion we assumed that $q_{2}$\ is scalar. The jump moments are given by
\begin{eqnarray}
q_{1}(x^{\prime }) &=&\int dx\ \left( x-x^{\prime }\right) \ q(x^{\prime },x)\\
	q_{2}(x^{\prime }) &=& \int dx\ \left\vert x-x^{\prime}\right\vert ^{2}q(x^{\prime },x).  \label{m3}
\end{eqnarray}

Substitution of Eq. \ref{m2} into Eq. \ref{m1} gives the Fokker-Planck equation\footnote{Alternative names are the Smoluchowski equation or the second Kolmogorov
equation.}
\begin{equation}
\frac{\partial f(x,t)}{\partial t}=-\frac{\partial }{\partial x}
\Big(q_{1}(x)f(x,t)\Big)+\frac{1}{2}\frac{\partial ^{2}}{\partial x^{2}}\Big(q_{2}(x)f(x,t)\Big).
\label{m4}
\end{equation}
It is now convenient to rename
\begin{equation}
D(x)\equiv \frac{1}{2}q_{2}(x)\text{ \ \ and \ \ }V(x)\equiv q_{1}(x)-\frac{1}{2}\frac{\partial }{\partial x}q_{2}(x).  \label{m5}
\end{equation}
where the diffusion coefficient $D$ is a matrix and the velocity $V$ a vector in phase space. The Fokker-Planck equation can then be written in the form
\begin{equation}
\frac{\partial f(x,t)}{\partial t}=-\frac{\partial }{\partial x}\cdot
J(x,t)
\end{equation} 
with 
\begin{equation}
  J(x,t)=V(x)f(x,t)-D(x)\frac{\partial }{\partial x}f(x,t),  \label{eqn-17}
\end{equation}
in which $J(x,t)$ is a probabilistic flux. Equations (\ref{eqn-16}) and (\ref{eqn-17}) give expressions for the flux in terms of the distribution $f(x,t)$.  The positive definite nature of the entropy production can be proven within the theory \cite{qqt-jsp,qian-jpc-02,qian-pre-04,qian-epjst}: 
\begin{eqnarray}
\frac{\mathrm{d}S}{\mathrm{d}t} &=&-k_{\text{B}}\frac{\mathrm{d}}{\mathrm{d}t}\int_{\Omega }f(x,t)\ln f(x,t)\mathrm{d}x  \nonumber \\
&=&k_{\text{B}}\oint_{\partial \Omega }\ln f(x,t)J(x,t)\cdot \mathrm{d}
\Omega -k_{\text{B}}\int_{\Omega }J(x,t)\cdot \frac{\partial }{\partial x}
\ln f(x,t)\mathrm{d}x  \nonumber \\[6pt]
&=&\frac{\mathrm{d}_{e}S}{\mathrm{d}t}+\frac{\mathrm{d}_{i}S}{\mathrm{d}t},
\label{eqn-18}
\end{eqnarray}
where $\oint_{\partial \Omega }$\ ...d$\Omega $ indicates an integral over the surface. The surface element d$\Omega $\ has the outward direction normal to the surface. The system is now open. Furthermore,
\begin{eqnarray}
\frac{\mathrm{d}_{i}S}{\mathrm{d}t} &=&k_{\text{B}}\int_{\Omega }J(x,t)\cdot \Big(D^{-1}(x)V(x)-\frac{\partial }{\partial x}\ln f(x,t)\Big)\mathrm{d}x\geq 0,  
\label{m6} \\
\frac{\mathrm{d}_{e}S}{\mathrm{d}t} &=&k_{\text{B}}\oint_{\partial \Omega
}\ln f(x,t)J(x,t)\cdot \mathrm{d}\Omega 
\nonumber\\
  && -k_{\text{B}}\int_{\Omega
}J(x,t)\cdot D^{-1}(x)V(x)\mathrm{d}x.  \label{m7}
\end{eqnarray}
The reason for the choices of $\frac{\mathrm{d}_{i}S}{\mathrm{d}t}$\ and $%
\frac{\mathrm{d}_{e}S}{\mathrm{d}t},$\ is that  $\frac{\mathrm{d}_{i}S}{\mathrm{d}t}$\ being consistent with
Onsager's idea of ``force $\times$ flux'', and it is non-negative.  This assures validity of the Second law of Thermodynamics, and it establishes a link to nonequilibrium thermodynamics. A complete parallel can be developed for the discrete description, Eq. (\ref{eqn-16}) \cite{schnakenberg}. The Markov theory, which is formulated in a phase space, has fluxes which are described by the single entity $J(x,t)$, the flux of probability density.

The current stochastic thermodynamics begins with the notion of entropy production in stochastic processes, used already in Hill's stochastic  cycle kinetics \cite{hill-chen-75,hill-book,schnakenberg} and  in Qians' work on irreversible Markov processes \cite{qq-79,qqq-81,qq-85,qqt-jsp,ge-qian-2010}.  Another origin are the fluctuation theorems and the Jarzynski-Crooks equality. See \cite{jarzyn,seifert-rrp,e-vdb,evdb_2010} for comprehensive reviews on the subject.

\section{Nonequilibrium Thermodynamics of Driven Cycles}

\label{sec:3}

Stochastic thermodynamics is a mesoscopic theory in terms of probability. One of the fundamental insights from the Hill's nonequilibrium thermodynamic theory \cite{hill-book} is the central role of {\it kinetic cycles}, both in steady state and in finite time. Actually, by realizing that entropy production is a fundamental property of each and every kinetic cycle, and that cycles are completed one by one stochastically in time \cite{hill-chen-75}, Hill and Chen indeed have implicitly conceived the notion of entropy production at the finite time \cite{qqq-81,qq-85}. It can be mathematically shown that the entropy production for a {\it stationary} Markov jump process with transition rates $q_{ij}$, e.g., systems following Eq. \ref{eqn-16}, has a cycle representation \cite{jqq}: 
\begin{eqnarray}
\frac{\mathrm{d}_{i}S}{\mathrm{d}t} &=&k_{\text{B}}\sum_{i>j}\Big(J_{ij}-J_{ji}\Big)\ln \left( \frac{J_{ij}}{J_{ji}}\right)  \label{m9} \\
&=&k_{\text{B}}\sum_{\text{all cycle }\Gamma }\Big(J_{\Gamma }^{+}-J_{\Gamma
}^{-}\Big)\ln \left( \frac{J_{\Gamma }^{+}}{J_{\Gamma }^{-}}\right) ,
\label{epr} \\
\ln \left( \frac{J_{\Gamma }^{+}}{J_{\Gamma }^{-}}\right) &=&\ln \left( 
\frac{q_{i_{0}i_{1}}q_{i_{1}i_{2}}\cdots q_{i_{n-1}i_{n}}q_{i_{n}i_{0}}}{q_{i_{0}i_{n}}q_{i_{n}i_{n-1}}\cdots q_{i_{2}i_{1}}q_{i_{1}i_{0}}}\right) ,
\label{sep}
\end{eqnarray}
in which $\Gamma $-cycle $=\big\{i_{0},i_{1},\cdots ,i_{n},i_{0}\big\}$, where all $i_{k}$ are distinct. Since a Markov process completes cycles stochastically, one can compute a finite-time entropy production along a stochastic trajectory by following the cycles. For a Markov process in equilibrium the detailed balance, $J_{ij}=J_{ji}$, is valid. The entropy
production for every cycle is then zero. The entropy production is the sum of the entropy productions of the separate cycles. The entropy production per cycle is $k_{\text{B}}\ln \big(J_{\Gamma}^{+}/J_{\Gamma }^{-}\big)$; the rate by which a particular cycle is being completed is $(J_{\Gamma }^{+}-J_{\Gamma }^{-})$; thus the entropy production {\it rate} per cycle is $k_{\text{B}}\big(J_{\Gamma }^{+}-J_{\Gamma }^{-} \big) \ln \big(J_{\Gamma}^{+}/J_{\Gamma }^{-}\big)$ \cite{jqq}. While computing
all the rates is challenging, it is amazing to observe that the entropy production per cycle  $k_{\text{B}}\ln \big(J_{\Gamma}^{+}/J_{\Gamma }^{-}\big)$,  is completely determined by the ratio of transition probability rates. This observation led Hill to suggest that kinetic cycles, not states, are fundamental units of NET.  

Cyclic processes, which were extensively investigated by Carnot, Clausius, Kelvin, and many others in the 19th century, can be described as thermodynamic processes in phase space. The beauty of the stochastic description is that the physical processes are all characterized by probabilities. In applications, however, the various flux terms can and should be interpreted as temperature driven, chemical-potential driven, or mechanically driven, etc.  We will now illustrate this by considering  simple examples.

Let us restrict the analysis to the  systems with discrete states. Following Esposito \cite{esposito-12}, we consider a mesoscopic system in state $i$ with internal energy $U_{i}$, entropy $S_{i}$, and number of particles $N_{i}$. If the mesoscopic system is completely isolated from its environment, then it will remain in the $i$ state indefinitely with conserved $U_{i},S_{i},N_{i}$. It has an equation of state $V=V_{i}(U_{i},S_{i},N_{i})$ where $V$ is the volume of the system. Now if the system is in contact with a heat bath with temperature $T$, and a material reservoir with chemical
potential $\mu $, then the state has a grand potential, also called Landau potential: 
\begin{equation}
\varphi _{i}\left( T,\mu \right) =U_{i}-TS_{i}-\mu N_{i}.  \label{landau-p}
\end{equation}
Transition to a state $j$ can occur due to the coupling to the heat and particle bath, with transition rates $q_{ij}$ and $q_{ji}$ which satisfy the detailed balance \cite{hill-book}: 
\begin{equation}
\frac{q_{ij}}{q_{ji}}=\frac{p_{j}^{eq}}{p_{i}^{eq}}=\exp \left( \frac{-\varphi _{j}+\varphi _{i}}{k_{B}T}\right). \label{ldb}
\end{equation}
Here we used the fact that in equilibrium the probability distribution is $p_{i,eq}\propto \exp \left( -\varphi _{i}/k_{B}T\right) $. Detailed balance implies that the forward and backward rates cannot be chosen independently if the potentials of the reservoirs are given. If one uses different heat and material reservoirs for the different states of the system, a stationary state may develop and obviously $p_{i}^{ss}q_{ij}\neq p_{j}^{ss}q_{ji}$ in that stationary state.

\subsection{Chemical cycle kinetics}

\label{sec3.1}

\begin{figure}[h]
\begin{center}
\includegraphics[width=3.2in]{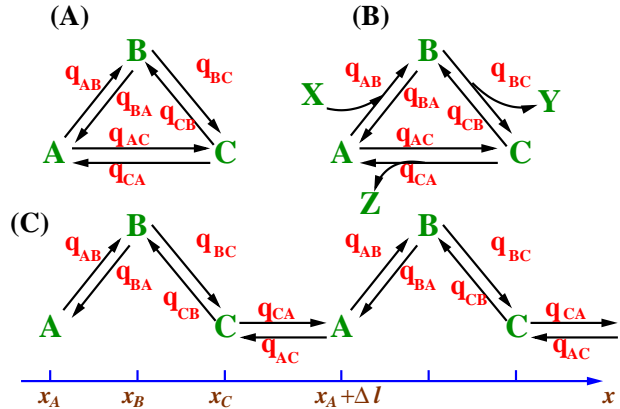}
\end{center}
\caption{(A) Three-state cycle kinetics in a closed system. (B) Three-state cycle kinetics in an open chemical system with material reservoirs of $X$, $Y $, and $Z$, with chemical potential $\mu ^{(X)}$, $\mu^{(Y)} $, and $\mu ^{(Z)}$. (C) Three-state cycle kinetics in an open chemical system with 1D spatial component $x$ (material reservoirs are not shown to avoid cluttering). A complete cycle kinetics accompanies a spatial displacement of $\Delta \ell $. Such a system has a \textquotedblleft tight\textquotedblright\ coupling between the cycle and the translocation. If there were nonzero transitions between $C(x_{C})$ and $A(x_{A})$, then the system would be loosely coupled  \cite{qian-mie}.}
\label{fig1}
\end{figure}

Consider a cycle as shown in Fig. \ref{fig1}(A), in which all three mesoscopic states $A$, $B$, and $C$ are in contact with the same heat bath with temperature $T$. Then, 
\begin{equation}
\hspace{-1cm} \frac{q_{AB}q_{BC}q_{CA}}{q_{BA}q_{CB}q_{AC}}  =\exp \left( \frac{-\varphi _{B}+\varphi _{A}}{k_{B}T}\right) \exp \left(\frac{-\varphi _{C}+\varphi _{B}}{k_{B}T}\right) \exp \left( \frac{-\varphi_{A}+\varphi _{C}}{k_{B}T}\right)=1.  \label{23}
\end{equation}
This is the detailed balance for the cycle.

If we consider a cycle  in the open system with material reservoirs having chemical potentials $\mu_{A}$, $\mu_{B}$ and $\mu_{C}$ [see Fig 1(B)], then potentials $\varphi _{i}$\ can be replaced by the corresponding chemical potentials $\mu _{i}$ as the reaction only changes $N_{i}$ in Eq. (\ref{landau-p}). Now, at constant $T$, if the first transition in the cycle is a part of the chemical reaction $A+X\rightleftharpoons B$, as shown in Fig. \ref{fig1}(B), and similarly the second transition is a part of $B \rightleftharpoons C+Y$, and the third transition involves $C \rightleftharpoons A+Z$, then one has the entropy production per cycle, or {\it cycle affinity} 
\begin{eqnarray}
{A}_{C} &=&k_{B}\ln \left( \frac{q_{AB}q_{BC}q_{CA}}{q_{BA}q_{CB}q_{AC}}\right)  \nonumber \\
&=&\left( \frac{\mu _{A}+\mu ^{(X)}-\mu _{B}}{T}\right) +\left( \frac{\mu_{B}-\mu _{C}-\mu ^{(Y)}}{T}\right) +\left( \frac{\big(\mu _{C}-\mu _{A}-\mu^{(Z)}}{T}\right)  \nonumber \\
&=&\frac{\mu ^{(X)}-\mu ^{(Y)}-\mu ^{(Z)}}{T}.  \label{e24}
\end{eqnarray}
Here the numerator  is the chemical potential difference associated with the reaction of \textquotedblleft external chemical potential reservoirs\textquotedblright\ $X \rightleftharpoons Y+Z$. This is an open chemical system with a chemical-potential driven cycle. The difference between state $B$ and state $A$ is one $X$, between states $B$ and $C$ is one $Y$, and between states $C$ and $A$ is one $Z$. The corresponding steady-state cycle flux is \cite{hill-book} 
\begin{equation}
J_{C}=\frac{q_{AB}q_{BC}q_{CA}-q_{AC}q_{CB}q_{BA}}{\left\{ \begin{array}{l}
q_{BC}q_{CA}+q_{CB}q_{BA}+q_{BA}q_{CA}+q_{CA}q_{AB}+ \\ 
q_{AC}q_{CB}+q_{CB}q_{AB}+q_{AB}q_{BC}+q_{BA}q_{AC}+q_{AC}q_{BC}
\end{array} \right\}}.  \label{3stateflux}
\end{equation}
It is clear that $J_{C}\times {A}_{C}\geq 0$ \cite{qian-jpc-06}. This is the entropy production in Eq. (\ref{epr}), for one cycle. It is zero if and only if Eq. (\ref{23}) holds true, i.e. if the chemical system is closed. It is zero for an open system if $\Delta \mu =\mu ^{(X}-\mu ^{(Y)}-\mu ^{(Z)}=0$, i.e., when the system is coupled to an equilibrium chemical bath.

In chemistry, it is often conveniently to write as
\begin{equation}
\frac{q_{AB}}{q_{BA}}=\exp \left( \frac{\mu _{A}+\mu ^{(X)}-\mu _{B}}{k_{B}T}\right) =\frac{q_{AB}^{o}a_{X}}{q_{BA}},
\end{equation}
in which $q_{AB}^{o}$ is a {\it second-order} rate constant, and $a_{X}=\exp \left( \mu ^{(X)}/k_{B}T\right) $ is the activity of species $X$. It follows that $q_{AB}^{o}$\ and $q_{BA}$\ satisfy the detailed balance,
\begin{equation}
\frac{q_{AB}^{o}}{q_{BA}}=\exp \left( \frac{\mu _{A}-\mu _{B}}{k_{B}T}
\right) .
\end{equation}
Recalling discussion in Section 2.3.1, we see that $q_{AB}^{o}$ and $q_{BA}$ are the original rate coefficients, which satisfy detailed balance, while $q_{AB}$ and $q_{BA}$ are the pseudo first-order rate coefficients, which do not.

\subsection{Chemomechanical cycle and a molecular motor}

Now if the mesoscopic system has a 1D position $x$ that experiences a constant external mechanical resistant force $\xi $ (or a rotational angle with a constant external torque) and undergoes cyclic motion, as shown in Fig. \ref{fig1}(C) \cite{hill,qian-bpc-97,fisher-2,fisher-1}  then Eq. (\ref{landau-p}) modifies into 
\begin{equation}
\varphi \left( T,\mu ;x\right) =U_{i}-TS_{i}-\mu N_{i}-\xi x,
\end{equation}
in which the term $\xi x$ should be compared with the $pV$ term in macroscopic thermodynamics. The entropy production per cycle, or cycle affinity, in Eq. (\ref{e24}) becomes \cite{qian-bpc-00} 
\begin{equation}
{A}_{C}=\frac{\mu ^{(X)}-\mu ^{(Y)}-\mu ^{(Z)}-\xi \Delta \ell }{T}.
\end{equation}
The significance of this result is that it establishes, mathematically, a mesoscopic free-energy balance between input Gibbs energy $\Delta \mu \equiv \mu ^{(X)}-\mu ^{(Y)}-\mu ^{(Z)}$, which becomes the work against the external force $\xi \Delta \ell $, and dissipation ${A}_{C}$, both per unit of flux. The efficiency of the chemomechanical energy transduction of the cycle immediately follows: $\eta _{\text{chemomechanic}}=\xi \Delta\ell /\Delta \mu $ \cite{qian-bpc-00}. One can also see that when the external force is given by $\xi =\Delta \mu /\Delta \ell $, known as a stalling force, the efficiency is 1; but at the same time the output mechanical power, e.g., the work per unit time, is zero. This is a
pathological consequence of assuming a single cycle that tightly couples the mechanical and chemical steps \cite{qian-mie}.  If this is not the  case, e.g., the chemical step and the mechanical step can ``slip'', then there will be at least one additional cycle in which  the chemical energy dissipates.

If the force $\xi $ is negative, it can push a negative $\Delta \mu $. Such a kinetic cycle will have mechanical force driven chemical pumping $Y+Z\longrightarrow X$, as in F$_{0}$F$_{1}$-ATP synthesis \cite{oster}.

\subsection{Temperature-driven kinetic cycle and thermomechanical efficiency}

\label{sec:3.3}

Let us again consider the cycle kinetics presented   Fig. \ref{fig1}(A). This time, the three mesoscopic states $A$, $B$, and $C$ are  in a contact with different temperature baths. Let us assume that $A,\ B,\ C$ have the temperatures $T_{A}$, $T_{B}$ and $T_{C}$. We further assume all chemical potentials are equal. Then, one has the cycle affinity given by \cite{esposito-epl,lervik-2} 
\begin{eqnarray}
{A}_{C} &=&k_{B}\ln \left( \frac{q_{AB}q_{BC}q_{CA}}{q_{BA}q_{CB}q_{AC}}\right) \\
&=&\left( \frac{\varphi _{A}-\varphi _{B}}{T_{A}}\right) +\left( \frac{\varphi _{B}-\varphi _{C}}{T_{B}}\right) +\left( \frac{\varphi _{C}-\varphi_{A}}{T_{C}}\right)  \nonumber \\
&=&\frac{U_{A}-U_{B}}{T_{A}}+\frac{U_{B}-U_{C}}{T_{B}}+\frac{U_{C}-U_{A}-\xi\Delta\ell}{T_{C}}  \nonumber \\
&=&U_{A}\left( \frac{1}{T_{B}}-\frac{1}{T_{C}}\right) +U_{B}\left( \frac{1}{T_{B}}-\frac{1}{T_{A}}\right) +U_{C}\left( \frac{1}{T_{C}}-\frac{1}{T_{B}}\right)-\frac{\xi\Delta\ell}{T_C}, \nonumber \\
\label{eq-29}
\end{eqnarray}
in which the force $\xi$ is a property of the external environment \cite{qian-bpc-97,fisher-1}. In the special case when $T_{B}=T_{C}$, one obtains 
\begin{eqnarray}
{A}_{C} &=&\frac{U_{A}-U_{B}}{T_{A}}+\frac{U_{B}-U_{A}-\xi \Delta\ell }{T_{B}}  \nonumber \\
&=&\left( U_{A}-U_{B}\right) \left( \frac{1}{T_{A}}-\frac{1}{T_{B}}\right) -
\frac{\xi \Delta \ell }{T_{B}}\ \geq 0,
\end{eqnarray}
Note, when $\xi\Delta\ell$ is positive,  $(T_{A}-T_{B})$ and $\mathcal{Q\equiv \ }(U_{A}-U_{B})$ always have opposite signs; thus the product $(U_A-U_B)\big(T_{A}^{-1}-T_{B}^{-1}\big)$ is  always positive.  Without loss of generality, we let $T_A>T_B$.  Then, the thermo-mechanical (first-law) efficiency can be defined as 
\begin{equation}
\eta _{\text{thermomechanic}}=\frac{\xi \Delta \ell }{|\mathcal{Q}|}=\frac{T_{B}}{\mathcal{Q}}{A}_{C}+\left( 1-\frac{T_{B}}{T_{A}}\right) \leq 1-\frac{T_{B}}{T_{A}}.  \label{c-limit}
\end{equation}
The maximal first-law efficiency\footnote{The term  {\it first-law efficiency} is used to distinguish it from the second-law efficiency (also known as a rational efficiency and exergy efficiency) which computes the efficiency of a process taking the  Second Law of Thermodynamics into account in practical engineering. The exergy of a system is the maximum useful work possible during a process  that brings the system into equilibrium with a heat bath.  Note that in chemomechanical energy transduction,  taking the Second Law into account does not reduce the  upper limit of efficiency $\eta_{\text{chemomechanical}}$,  when the power is zero.}
is the Carnot limit \cite{esposito-epl}. The second-law efficiency  is then equal to \cite{honerkamp}:
\begin{equation}
\eta _{\text{exergy}}=\frac{\xi \Delta \ell }{\xi \Delta \ell +T_{B}%
{A}_{C}}=\frac{\xi \Delta \ell }{|\mathcal{Q}|\left( 1-\frac{T_{B}}{%
T_{A}}\right) }.
\end{equation}

The entropy production is the product of $J_{C}$, given in Eq. (\ref{3stateflux}), and ${A}_{C}$. With given high and low temperature baths $T_{A}$ and $T_{B}$, one could ask a different question. Allowing temperature $T_{C}$ to be between $T_{A}$ and $T_{B}$, what is the condition for the maximum power for a given entropy production? The answer is that this situation is realized when
\begin{equation}
\frac{q_{AB}}{q_{BA}}=\frac{q_{BC}}{q_{CB}}=\frac{q_{CA}}{q_{AC}},
\label{maxpower}
\end{equation}
Eq. (\ref{maxpower}) is known as the \textquotedblleft principle of constant force\textquotedblright in the field of molecular motors \cite{oster,qian-jmc}. It also corresponds to  equal chemical potential drops in the  metabolic engineering \cite{timing}. In this case, one can write
\begin{equation}
k_{\text{B}}\ln \left( \frac{q_{AB}}{q_{BA}}\right) =k_{\text{B}}\ln \left( \frac{q_{BC}}{q_{CB}}\right) =k_{\text{B}}\ln \left( \frac{q_{CA}}{q_{AC}}\right),
\end{equation}
which is a principle of constant entropy production. Eq. (\ref{maxpower}) is called the principle of constant thermodynamic force in nonequilibrium thermodynamics.  It is interesting to note that the same  result has been found to characterize stationary-state operation of process units at minimum entropy production. 
\cite{johannessen-kjelstrup-05}

\subsection{Non-isothermal enzyme kinetic cycle}

Sec. \ref{sec:3.3} illustrated the thermomechanical energy transduction and obtained the Carnot efficiency. Nonequilibrium chemical or biochemical cycles can also be induced by temperature difference, and vice versa; thermochemical coupling can occur in an enzyme that operates under non-isothermal environment. Indeed, intracellular enzyme mediated biochemical reactions \emph{in situ} are usually chemical-potential driven NESS cycles \cite{qian-bj,xie}. Michaelis-Menten-Briggs-Haldane kinetics of an individual enzyme, with a single substrate and a single product, can be best understood as a steady state flux $J_{C}=J_{C}^{+}-J_{C}^{-}$ \cite{hill-book,kjelstrup-qian,kolomeisky-book} of the kinetic cycle in Fig. \ref{fig1}B without the $Y$, with a single temperature $T$. The one-way cycle fluxes $J_{C}^{\pm
}$ are probability weighted inverse of the mean first-passage time \cite{qian-bj,qian-xie,kolomeisky-book}. We will identify $X$ and $Z$ as the substrate $S$ and the product $P$ of the enzyme, with $q_{AB}=q_{AB}^{o}a_{S}$ and $q_{AC}=q_{AC}^{o}a_{P}$ where 
\begin{equation}
\frac{q_{AB}^{o}}{q_{BA}}=\exp \left( \frac{\varphi _{B}-\varphi _{A}}{k_{B}T}\right) ,\ \mu_{S}\equiv \mu^{o}_{S}+k_{B}T\ln a_{S},
\end{equation}
\begin{equation}
\frac{q_{CA}}{q_{AC}^{o}}=\exp \left( \frac{\varphi _{C}-\varphi _{A}}{k_{B}T}\right) ,\ \mu_{P}\equiv \mu^o_{P}+k_{B}T\ln a_{P},
\end{equation}
where $\varphi _{A},\varphi _{B},\varphi _{C}$ are Landau potentials, given in Eq. (\ref{landau-p}), at temperature $T$, and $a_{S}$ and $a_{P}$ are the dimensionless chemical activities of the substrate and the product. For sufficiently dilute solution, they are the same as the molecular concentrations $c_{S}$ and $c_{P}$, divided by the standard concentration, $c_{0}=1$ mole/L.  To be consistent with the notions in the biochemical literature, we will assume that the solution is always ideal. Then Eq. (\ref{3stateflux}) becomes 
\begin{equation}\label{e50}
J_{C}=\frac{\left( \frac{V_{max}^{f}}{K_{M}^{f}}\right) c_{S}-\left( \frac{V_{max}^{b}}{K_{M}^{b}}\right) c_{P}}{1+\frac{c_{S}}{K_{M}^{f}}+\frac{c_{P}}{K_{M}^{b}}},
\end{equation}
in which  Michaelis constants and maximal velocities of the forward and backward reactions, with corresponding $J_{C}^{+}$ and $J_{C}^{-}$, are equal to
\begin{eqnarray}
K_{M}^{f} &=&\frac{q_{BC}q_{CA}+q_{CB}q_{BA}+q_{BA}q_{CA}}{q_{CA}q_{AB}^{o}+q_{CB}q_{AB}^{o}+q_{AB}^{o}q_{BC}}, \\
V_{max}^{f} &=&\frac{q_{AB}^{o}q_{BC}q_{CA}}{q_{CA}q_{AB}^{o}+q_{CB}q_{AB}^{o}+q_{AB}^{o}q_{BC}}, \\
K_{M}^{b} &=&\frac{q_{BC}q_{CA}+q_{CB}q_{BA}+q_{BA}q_{CA}}{q_{AC}^{o}q_{CB}+q_{BA}q_{AC}^{o}+q_{AC}^{o}q_{BC}}, \\
V_{max}^{b} &=&\frac{q_{AC}^{o}q_{CB}q_{BA}}{q_{AC}^{o}q_{CB}+q_{BA}q_{AC}^{o}+q_{AC}^{o}q_{BC}}.
\end{eqnarray}
One can find these complicated expressions  in standard enzyme kinetics texts, e.g., \cite{segal}. When $B\rightarrow C$ is a rate-limiting step and $q_{BC}$ and $q_{CB}$\ are much smaller than the others, one has $K_{M}^{f}=q_{BA}/q_{AB}^{o}$, which is the original Michaelis constant.

These equations can be viewed also a statement about the cycle affinity in Eq. (\ref{e24}) \cite{qian-xie} 
\begin{equation}
\frac{\left( \frac{V_{max}^{f}}{K_{M}^{f}}\right) }{\left( \frac{V_{max}^{b}}{K_{M}^{b}}\right) }=\exp \left( \frac{{A}_{C}}{k_{B}}\right)
=e^{\left( \mu ^{(S)}-\mu ^{(P)}\right) /k_{B}T}.
\end{equation}
The NESS entropy production then is $J_{C}\times {A}_{C}$: the number of cycles completed per unit time $\times$ the entropy production per cycle.

We now consider a non-isothermal situation as in \cite{kjelstrup-qian}: The enzyme is assumed to reside in a membrane with a temperature $T^{(1)}$. It separates two bulk solutions both with a temperature $T^{(2)}\neq T^{(1)}$. We can then generalize the enzyme kinetics to non-isothermal condition with $BC$ transitions under $T^{(1)}$ and the other two under $T^{(2)}$. Such an enzyme kinetic cycle is simply a thermochemical system, a special case of the mesoscopic thermo-chemo-mechanical machine. Then we have the cycle affinity given by
\begin{eqnarray}
{A}_{C} &=&\frac{U_{A}+\mu ^{(S)}-U_{B}}{T^{(2)}}+\frac{U_{B}-U_{C}}{T^{(1)}}+\frac{U_{C}-U_{A}-\mu ^{(P)}}{T^{(2)}}  \nonumber \\
&=&\Big(U_{B}-U_{C}\Big)\left( \frac{1}{T^{(1)}}-\frac{1}{T^{(2)}}\right) +\frac{\mu ^{(S)}-\mu ^{(P)}}{T^{(2)}}  \nonumber \\
&=&\mathcal{Q}^{\text{(measurable heat)}}\left( \frac{1}{T^{(1)}}-\frac{1}{T^{(2)}}\right) +{A}_{C}^{\text{(substrate turnover)}}.
\end{eqnarray}
The function $\mathcal{Q}^{\text{(measurable heat)}}$ is different from the total heat which should contain the part of energetic change in $\mu ^{(S)}-\mu ^{(P)}$ \cite{bk-08}. At the same time, since the transitions between $B$ and $C$ are under $T^{(1)}$ and the other transitions are under $T^{(2)}$, the NESS cycle flux $J_{C}$ in Eq. (\ref{e50}) can be expressed as \begin{eqnarray}
&&J_{C}=\frac{q_{AB}q_{CA}\left( \frac{q_{BC}}{q_{CB}}\right) -q_{AC}q_{BA}}{(q_{CA}+q_{AB}+q_{AC})\frac{q_{BC}}{q_{CB}}+q_{BA}+q_{AC}+q_{AB}+\frac{q_{BA}q_{CA}+q_{CA}q_{AB}+q_{BA}q_{AC}}{q_{CB}}}  \nonumber \\
&=&\frac{\left[ \left( \frac{V_{max}^{f}}{K_{M}^{f}}\right) c_{S}-\left( 
\frac{V_{max}^{b}}{K_{M}^{b}}\right) c_{P}\right] ^{T=T^{(2)}}}{1+\frac{c_{S}}{K_{M}^{f}}+\frac{c_{P}}{K_{M}^{b}}}\left\{ 1+\frac{\exp \left[ \frac{-U_{B}+U_{C}}{k_{B}}\left( \frac{1}{T^{(1)}}-\frac{1}{T^{(2)}}\right) \right]
-1}{1-\exp \left( \frac{\mu ^{(P)}-\mu ^{(S)}}{k_{B}T^{(2)}}\right) }\right\} .  \nonumber \\
&&
\end{eqnarray}
The last term in the curly brackets is a correction term for Michaelis-Menten kinetics due to non-isothermal condition. In the linear regime, the temperature difference-driven catalytic flux is equal to
\begin{equation}
\left[ \frac{\left( \frac{V_{max}^{f}}{K_{M}^{f}}\right) c_{S}}{1+\frac{a_{S}}{K_{M}^{f}}+\frac{a_{P}}{K_{M}^{b}}}\right] ^{eq}\left\{ \frac{U_{C}-U_{B}}{k_{B}}\left( \frac{1}{T^{(1)}}-\frac{1}{T^{(2)}}\right) \right\} ,
\end{equation}
and the chemical potential difference-driven heat flux is 
\begin{equation}
\left[ \frac{\left( \frac{V_{max}^{f}}{K_{M}^{f}}\right) c_{S}}{1+\frac{c_{S}}{K_{M}^{f}}+\frac{c_{P}}{K_{M}^{b}}}\right] ^{eq}\left\{ \frac{\mu
^{(S)}-\mu ^{(P)}}{k_{B}T^{(2)}}\right\} .
\end{equation}
They have the same coefficient $[\cdots ]^{eq}$, the one-way flux in equilibrium, as expected from the Hill's theory.

\subsection{Chemical-potential driven enzyme selectivity amplification}

There is a very interesting example for the application of mesoscopic NET. It concerns with regulations  of intracellular communication signals in terms of enzyme  activities.	 Enzymes found in the living organisms has specific interaction with its cognate substrate molecules. The notion of biochemical specificity between an enzyme and its substrate has been quantified, traditionally, in terms of their equilibrium association constant.  Therefore, an enzyme $E$ interacting with two different substrates, one cognate $X$  and another noncognate $Y$, via the following chemical reactions
\begin{equation*}
E+X \rightleftharpoons EX,\quad E+Y \rightleftharpoons EY,
\end{equation*}
with respective equilibrium association constants $K_{X}$ and $K_{Y}$, is expected to have the selectivity for $X$ over $Y$ given by the ratio $K_X/K_Y$.   Recall that $K_{X}=[EX]^{eq}\left( [E]^{eq}[X]^{eq}\right) ^{-1}$ and $K_{Y}=[EY]^{eq}\left( [E]^{eq}[Y]^{eq}\right) ^{-1}$. The function $K_X/K_Y$, thus, is equal to the ratio between the equilibrium concentrations $[EX]$ and $[EY]$,  when there is an equal amount of $X$ and $Y$. However, it has been discovered that in  living cells, the selectivity of an enzyme toward its cognate substrate can be much greater than the $K_X/K_Y$.   This phenomenon has been termed {\it selectivity amplification}. These deviations in selectivity are clearly connected to the nonequilibrium nature of biological processes in living cells.

One can also recognize  the ratio $K_{X}/K_{Y}$ as the equilibrium constant for the {\it ligand exchange reaction} \cite{qian-jmb-06} 
\begin{equation}
EY+X \rightleftharpoons EX+Y.  \label{eq-47-rxn}
\end{equation}
It is important to note that both $K_{X}$ and $K_{Y}$ are determined by the molecular structures and interactions between the enzyme and the ligands, which  depend on the temperature, pH and solvents.

\begin{figure}[h]
\begin{center}
\includegraphics[width=3.2in]{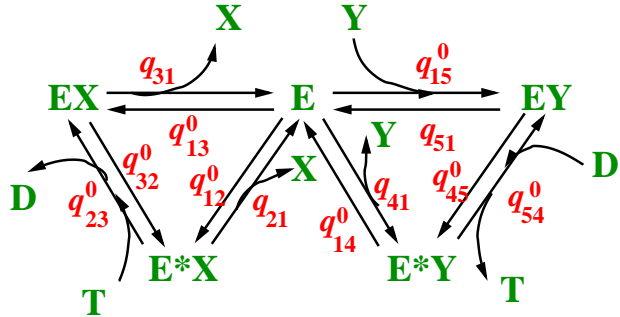}
\end{center}
\caption{ Cycle kinetics driven by the chemical potential difference between  $T$ and $D$, $\Delta \protect\mu _{TD}$. There are two kinetic cycles going through states $E\rightarrow E^{\ast }X\rightarrow EX\rightarrow E$, on the left, and $E\rightarrow E^{\ast }Y\rightarrow EY\rightarrow E$, on the right. }
\label{fig3}
\end{figure}

A fundamental role in biology is played by the concept  of {\it kinetic proofreading}. It is  a  mechanism for altered selectivity, which  uses driven biochemical reactions with fluxes to regulate enzymatic specificity, breaking the conventional wisdom that enzymatic specificity is defined solely by the equilibrium affinity. More specifically, it places the reaction in Eq. (\ref{eq-47-rxn}) inside a driven kinetic cycle such that the ratio of concentrations is given by 
\begin{equation}
\theta \equiv \frac{\lbrack EX][Y]}{[EY][X]}=\frac{\left( \frac{[EX]}{[E][X]}
\right) }{\left( \frac{[EY]}{[E][Y]}\right) }.
\end{equation}
In a driven NESS $\theta $ can be significantly different from its equilibrium value $K_{X}/K_{Y}$.

To understand the kinetic proofreading let us consider Fig. \ref{fig3} that shows a kinetic scheme in which the association-dissociation reaction is coupled to a reaction $T \rightleftharpoons D$. Then, when $T$ and $D$ are not in their chemical equilibrium, there will be two kinetic cycles: one couples the $E+X \rightleftharpoons EX$ with $T \rightleftharpoons D$, and the second one couples $E+Y \rightleftharpoons EY$ with $T \rightleftharpoons D$. The ratio of NESS concentrations can be computed, leading to 
\begin{eqnarray}
\theta &=&\left( \frac{[EX][Y]}{[EY][X]}\right) ^{NESS}\ =\ \frac{\displaystyle\left( \frac{q_{13}^{o}q_{21}+q_{13}^{o}q_{23}^{o}[T]+q_{23}^{o}q_{12}^{o}[T]}{q_{23}^{o}[T]q_{31}+q_{31}q_{21}+q_{21}q_{32}^{o}[D]}\right) }{\displaystyle \left( \frac{q_{15}^{o}q_{41}+q_{15}^{o}q_{45}^{o}[T]+q_{45}^{o}q_{14}^{o}[T]}{q_{45}^{o}[T]q_{51}+q_{51}q_{41}+q_{41}q_{54}^{o}[D]}\right) }
\label{eq49} \\[6pt]
&=&\frac{K_{X}}{K_{Y}}\left( \frac{q_{21}+q_{23}^{o}[T]+q_{23}^{o}q_{12}^{o}[T]/q_{13}^{o}}{q_{21}+q_{23}^{o}[T]+q_{23}^{o}q_{12}^{o}[T]/\gamma q_{13}^{o}}\right)
\left( \frac{q_{41}+q_{45}^{o}[T]+q_{45}^{o}q_{14}^{o}[T]/\gamma q_{15}^{o}}{q_{41}+q_{45}^{o}[T]+q_{45}^{o}q_{14}^{o}[T]/q_{15}^{o}}\right) ,  \nonumber
\end{eqnarray}
in which $k_{B}T\ln \gamma =\Delta \mu _{TD}$ $=\frac{q_{12}^{o}q_{23}^{o}q_{31}[T]}{q_{21}q_{32}^{o}q_{13}^{o}[D]}$ is the nonequilibrium thermodynamic force. The superscript $^{o}$ denotes second-order rate constants as indicated in Fig. \ref{fig3}.

When $T$ and $D$ have their equilibrium value, the detailed balance is satisfied and we have 
\begin{equation}
q_{12}^{o}q_{23}^{o}q_{31}/(q_{21}q_{32}^{o}q_{13}^{o})=([D]/[T])^{eq}=q_{14}^{o}q_{45}^{o}q_{51}/(q_{41}q_{54}^{o}q_{15}^{o}).
\end{equation}
Then 
\begin{equation}
([EX][Y]/[EY][X])^{eq}=([EX]/[E][X])^{eq}([E][Y]/[EY])^{eq}=K_{X}/K_{Y}.
\end{equation}
But for the deviations from equilibrium one obtains
\begin{equation}
\hspace{-2cm} \gamma \equiv q_{12}^{o}q_{23}^{o}q_{31}[T]/(q_{21}q_{32}^{o}q_{13}^{o}[D])=q_{14}^{o}q_{45}^{o}q_{51}[T]/(q_{41}q_{54}^{o}q_{15}^{o}[D])=e^{\Delta \mu _{TD}/k_{B}T}>1.
\end{equation}
In a well designed NESS biochemical network, $([EX][Y]/[EY][X])^{NESS}$ can be as high as $\gamma (K_{X}/K_{Y})$ and as low as $\gamma^{-1}(K_{X}/K_{Y}) $ \cite{qian_arpc}. 

Now if the enzyme has 
\begin{equation}
 \left( \frac{q_{12}^{o}}{\gamma q_{13}^{o}}\right) \ll \frac{q_{21}}{q_{23}^{o}[T]}\ll \left( \frac{q_{12}^{o}}{q_{13}^{o}}\right) ,
\end{equation}
and furthermore one assumes that the corresponding rate constants in the two kinetic cycles for $X$ and $Y$ are essentially the same except $q_{41}/q_{21}=q_{51}/q_{31}=K_{X}/K_{Y}$,  Hopfield and Ninio discovered  the mechanism of the high-fidelity protein biosynthesis \cite{jjh,ninio}. In this case,
\begin{equation}
\theta =\frac{K_{X}}{K_{Y}}\left( \frac{q_{23}^{o}q_{12}^{o}[T]/q_{13}^{o}}{q_{21}}\right) \left( \frac{q_{41}}{q_{45}^{o}q_{14}^{o}[T]/q_{15}^{o}}\right) =\left( \frac{K_{X}}{K_{Y}}\right) ^{2}.
\end{equation}
Obviously, the enzyme selectivity can be very different from the equilibrium estimates, and this is the essence of the kinetic proofreading mechanism.

\section{Coupling between Nonequilibrium Processes via Kinetic Cycles}

One of the most important new features that arise in nonequilibrium thermodynamics is the coupling terms between two types of transport processes \cite{onsager-31,onsager-31b,degroot-mazur, bedeaux-kjelstrup}. Onsager's pioneering work  elucidated a symmetry among the coupling coefficients in the force-flux relations when a system is near its equilibrium, which necessarily has a time-reversal symmetry \cite{onsager-31,onsager-31b}. In the context of mesoscopic chemical kinetics it was shown by Hill  \cite{hill-nature,hill-pnas-83}, that the Onsager coefficients can be expressed in terms of all  equilibrium one-way cycle fluxes that couple any two processes. The beauty of stochastic thermodynamics is that
the notion of \textquotedblleft coupling\textquotedblright\ can be formulated in phase space in terms of probabilistic fluxes irrespective of the microscopic details of underlying  physical and chemical processes.

\subsection{Cycles and the Onsager coefficients}

Hill's theory of the Onsager's reciprocal relation is based on kinetic cycles in discrete-state space, and it employs  a graph-theoretical treatment. Specifically, consider a irreducible Markov  process with $q_{ij}=0$ with all individual transition being reversible.  Let  the Markov network has $N$ non-zero reversible transitions, $e_{1},e_{2},\cdots ,e_{N},$ where $e$ stands for \textquotedblleft edge\textquotedblright , and $\kappa $ reversible cycles $c_{1},c_{2},\cdots ,c_{\kappa },$ where $c$ stands for \textquotedblleft cycle \textquotedblright. We give every transition and cycle a defined direction, and denote a set ${E}=\{e_{1},e_{2},\cdots ,e_{N}\}$ be the set of all  transitions with nonzero net flux in NESS. Similarly, ${C}=\{c_{1},c_{2},\cdots ,c_{\kappa }\}$ is the set of all the cycles with nonzero net cycle flux.  Combinatorial calculations show that $N \leq {\frac{m!}{2!(m-2)!}}$ and $\kappa \leq \sum_{\ell =3}^{N}\frac{N!}{(N-\ell )!\ell }$. 

Now we can introduce an $N\times \kappa $ edge-to-cycle incidence matrix: 
\begin{equation}
\Theta _{ij}=\left\{ 
\begin{array}{cl}
+1 & \text{ if }e_{i}\in {E} \text{ is a step of }c_{j}\in {C}
,e_{i}\text{ and }c_{j}\text{ in same direction}; \\ 
-1 & \text{ if }e_{i}\in  {E} \text{ is a step of }c_{j}\in {C}
,e_{i}\text{ and }c_{j}\text{ in opposite direction}; \\ 
0 & \text{ if }e_{i}\in  {E} \text{ is not any step of }c_{j}\in {C}.
\end{array}
\right.
\end{equation}
For the same graph, there is also a $m\times N$  matrix $\Xi$, representing the signed incidence between node (state) to directed-edge (reversible transition).  Then each column of $\Theta $, a cycle, corresponds to a vector in the right null space of $\Xi $.

We can show that
\begin{equation}
J_{e_{i}}=\sum_{k,c_{k}\in {C}} \Theta _{ik}\Big(J_{c_{k}}^{+}+J_{c_{k}}^{-}\Big)\tanh \left( \frac{1}{2k_{B}}\sum_{\ell
,e_{\ell } \in{E}}\frac{\Theta _{\ell k}\Delta \mu _{e_{\ell}}^{(\nu )}}{T^{(\nu )}}\right),  \label{eqn3167}
\end{equation}
where $\Delta \mu _{e_{\ell }}^{(\nu )}$ is the chemical potential difference of transition $e_{\ell }\in {E}$ with the temperature $T^{(\nu )}$, $J_{c_{k}}^{+}$ and $J_{c_{k}}^{-}$ are the two opposite one-way cycle fluxes of the cycle $c_{j}\in {C}$ \cite{hill-book,qqq-81,jqq}.

When $|\Delta\mu^{(\nu)}_{e_{\ell}}|\ll k_BT^{(\nu)}$ is sufficiently small,  $(J^+_{c_k}+J^-_{c_k})\simeq 2J_{c_k}^{+,eq}$, and Eq. (\ref{eqn3167}) becomes 
\begin{equation}
J_{e_i} \simeq \sum_{\ell,e_{\ell}\in{E}} \left [ \sum_{k,c_k\in{C}} \Theta_{ik} J_{c_k}^{+,eq} \Theta_{\ell k} \right ] \left(\frac{\Delta\mu^{(\nu)}_{e_{\ell}}}{k_BT^{(\nu)}} \right).  \label{eqn-032}
\end{equation}
The term inside the square bracket  is a symmetric matrix $M_{i\ell}=M_{\ell i}$. Onsager's reciprocal relation is immediately observed. Hill called Eq. (\ref{eqn-032}) {\it the statistical mechanics of Onsager's principle} \cite{hill-nature}. Every kinetic cycle that links transitions $e_i$ and $e_{\ell} $ contributes to their coupling \cite{lervik}. One can in fact introduce a \emph{coupling efficiency} as $M_{i\ell}/\sqrt{M_{ii}M_{\ell\ell}}$.

For a single cycle with $N$ transitions, $\Theta $ is $N\times 1$ with all
elements 1, 
\begin{equation}
J_{e_{i}}=\big(J_{c}^{+}+J_{c}^{-}\big)\tanh \left( \frac{1}{2k_{B}}\sum_{\ell =1}^{N}\frac{\Delta \mu _{e_{\ell }}^{(\nu )}}{T^{(\nu )}}\right).
\end{equation}

\subsection{Kinematics and NET of Markov processes}

In nonequilibrium thermodynamics, a passive transport typically involves a constituent following its chemical potential difference. Active transport, on the other hand, typically involves the motion of a constituent \emph{against} its chemical potential. This cannot occur by itself. It needs the help of another process \cite{bk-08}. This is known as \textquotedblleft pumping \textquotedblright\ in classical mechanics and in biophysics. One of the most famous such examples  is P. D. Mitchell's chemiosmotic mechanism of ATP synthesis  in mitochondria of living cells \cite{mitchell}.

In stochastic thermodynamics, in the passive process a system will move from a state of low probability to a state of higher probability. Being able to include the temperature differences as driving forces for transport is a particular challenge for stochastic thermodynamics, which is based on the description of Markov dynamics. The result in Sec. \ref{sec:3.3} suggests that Hill's cycle kinetic approach is not merely a kinetic theory, but also a thermodynamic one \cite{hill-book}. In classical chemical thermodynamics, $\ln (q_{ij}/q_{ji})$ is an equilibrium thermodynamic quantity. It is now quite clear from the cycle representation of the steady-state entropy production, given in Eq. (\ref{epr}), that the term $\big(J_{\Gamma}^{+}-J_{\Gamma }^{-}\big)$ is the kinematics of a Markov process, while the term $\ln \big(J_{\Gamma }^{+}/J_{\Gamma }^{-}\big)$ contains the essential information of nonequilibrium thermodynamics of an individual cycle.

\section{Discussion and Future Directions}

The entropy balance equation (\ref{1}) is valid for a large number of nonequilibrium systems with \emph{phenomenological laws} describing irreversible, transport processes in the form of proportionalities, e.g., Fourier's law between heat flow and temperature gradient, Fick's law between flow of a component in a mixture and its concentration gradient, Ohm's law between electrical current and potential gradient, Newton's law between shearing force and velocity gradient, the law of mass action between reaction rate and chemical potentials \cite{degroot-mazur}. Each of these phenomena involves a \textquotedblleft flux\textquotedblright\ that characterizes transport of certain entities, like mass, charge or energy, in response to a thermodynamic force \cite{onsager-31,onsager-31b}.

Starting with Boltzmann's notion of entropy of a classical mechanical system with conserved mechanical energy $U$, fixed volume $V$, and number of particles $\{N_{k}\}$, the entropy $S\big(U,V,\{N_{k}\}\big)$ can be calculated using the microcanonical ensemble given the Hamiltonian. The Gibbs equation can be written in the form 
\begin{equation}
\mathrm{d}S=\frac{1}{T(U,V,\{N_{k}\})}\mathrm{d}U+\frac{p(U,V,\{N_{k}\})}{T(U,V,\{N_{k}\})}\mathrm{d}V-\sum_{k}\frac{\mu _{k}(U,V,\{N_{k}\})}{T(U,V,\{N_{k}\})}\mathrm{d}N_{k}.
\end{equation}
It is clear from the work of Boltzmann and Gibbs that a probability measure is needed to define the entropy. As was also clearly known to Boltzmann, there is simply no entropy production in a purely deterministic treatment of classical, smooth motions \cite{dorfman}. The Gibbs equation can be used to find both the flux and the production of entropy in a transport phenomenon \cite{degroot-mazur}. 

\subsection{The nature of stochastic dynamics}

The notion of entropic force is sometimes considered to be difficult. As observed by de Groot and Mazur: \textquotedblleft [E]ntropic forces have nothing to do with forces in the Newtonian sense \textquotedblright, and \textquotedblleft [P]erhaps the name {\em affinity} would have been preferable\textquotedblright \cite{degroot-mazur}.

In the theory of stochastic processes there is a universal equation of motion with probability fluxes in phase space. The present paper shows that starting from such a mesoscopic description, a complete NET, with fluctuations, can be developed.

There is a need for a conceptual clarification on the mathematical method of stochastic processes in the theory of mesoscopic NET. Kolmogorov's mathematical theory of stochastic processes \cite{vankampen} articulates a
logic separation between the abstract probability of \textquotedblleft random events\textquotedblright\ in a probability space, and random variables defined on the space\footnote{According to Kolmogorov, a probability space is a measure space, and random variables are measurable functions defined on the measure space.} as physical observables. Markov dynamics described by a probability function $f(x,t)$ follows a linear master equation. A theory of entropy and entropy
production, according to current stochastic thermodynamics, can be formulated at this abstract level in terms of probabilistic flux that devoids the specific nature of the underlying dynamic phenomena.

With this new found perspective, it becomes clear that the local equilibrium assumption has to be made only when applying stochastic thermodynamics to a system with observables, as was illustrated in Sec. \ref{sec:3}.

\subsection{The nature of nonequilibrium processes}

Classical thermodynamics is a theory about the emergent behavior of a macroscopic system. It is insensitive to the details of the equations of motions of individual particles within the system. In terms of the mesoscopic description of a system, nonequilibrium thermodynamics is a theory about emergent probabilistic behavior, and it is expected to be insensitive to the details of stochastic Markov dynamics.

The term \textquotedblleft nonequilibrium\textquotedblright\ deserves a clarification. To some authors, the notion of \textquotedblleft equilibrium\textquotedblright\ is a mechanical concept. Thus, according to this usage, an oscillatory Hamiltonian dynamics is non-equilibrium. To others, however, equilibrium is a statistical thermodynamic concept. There
are fluctuations in an equilibrium system. In the present work, we have used the term nonequilibrium in the statistical thermodynamic sense, as it most frequently utilized in Chemistry. Nonequilibrium processes lead to \textquotedblleft
irreversible\textquotedblright\ and \textquotedblleft dissipative\textquotedblright\ behavior. It can be quantified by a positive definite entropy production.

Nonequilibrium thermodynamics (NET), therefore, describes dynamic processes with dissipation. In a mesoscopic perspective in probabilistic terms, stationary transport phenomena concern with the cycle kinetics, cycle affinities, and cycle fluxes. The cycle affinity as a physical quantity is actually easy to compute. The complexities of NET are in the decomposition of a system into cycles and the computation of the cycle fluxes. A cycle flux, however, is \textquotedblleft driven\textquotedblright\ by thermodynamic forces. The detailed mesoscopic cycles, each with its own probability, and
their coupling to outside sources, yield the reciprocal relation first formulated by Onsager \cite{onsager-31,onsager-31b}.

It will be important to apply and to extend the presented here stochastic mesoscopic framework of NET for different chemical, physical and biological processes. This will help to clarify mechanisms of various complex phenomena from fundamental point of views.  

\section*{Acknowledgments}

We thank the Lorentz Center of Leiden University for hosting the workshop on \textquotedblleft Nanothermodynamics: For
Equilibrium and Nonequilibrium\textquotedblright\ (December 1-5, 2014) where the authors started this work. ABK was also supported by the Welch Foundation (Grant C-1559), by the NSF (Grant CHE-1360979), and by the Center for Theoretical Biological Physics sponsored by the NSF (Grant PHY-1427654).

\end{document}